\documentclass[aps, prl, reprint, twocolumn, amsmath, amssymb]{revtex4-1}
\usepackage[utf8]{inputenc}
\usepackage{graphicx}
\usepackage{bm}

\newcommand{\Cs}{\ensuremath{C_\mathrm{s}}}

\renewcommand{\vec}[1]{{\mathbf{#1}}}

\renewcommand{\d}{\mathrm{d}}

\newcommand{\dV}{\mathrm{dA}\,}

\newcommand{\bhat}{\hat{\vec b}}

\begin{document}
\title{Unified transport scaling laws for plasma blobs and depletions}
\author{M.\ Wiesenberger}
\email[E-mail: ]{Matthias.Wiesenberger@uibk.ac.at}
\affiliation{Institute for Ion Physics and Applied Physics, 
                     Universit\"at Innsbruck, A-6020 Innsbruck, Austria}
\author{M.\ Held}
\affiliation{Institute for Ion Physics and Applied Physics, 
                     Universit\"at Innsbruck, A-6020 Innsbruck, Austria}
\author{R.\ Kube}
\author{O.\ E.\ Garcia}
\affiliation{Department of Physics and Technology, UiT The Arctic University of Norway, N-9037 Tromsø, Norway}

\date{\today}

\begin{abstract} 
  We study the dynamics of seeded plasma blobs and depletions in an (effective) gravitational field.
  For incompressible flows 
  the radial center of mass velocity of blobs and depletions 
  is proportional to the
  square root of their initial cross-field size and amplitude. 
  If the flows are compressible, this scaling holds only
  for ratios of amplitude to size larger than a critical value.
  Otherwise, the maximum blob and depletion velocity depends 
  linearly on the initial amplitude and is independent of size. 
  In both cases the acceleration of blobs and depletions depends 
  on their initial amplitude relative to the background plasma density, 
  is proportional to gravity and independent of their cross-field size.
  Due to their reduced inertia plasma depletions accelerate more quickly 
  than the corresponding blobs. 
  These scaling laws are derived from the invariants of the governing
  drift-fluid equations and 
  agree excellently with numerical simulations over five orders
  of magnitude. 
  We suggest an empirical model that unifies and correctly captures the 
  radial acceleration and maximum velocities of both blobs and depletions.
\end{abstract}

\maketitle
Fluctuation induced transport across magnetic field lines is ubiquitous in
magnetized plasmas in various conditions.  In the scrape-off layer of tokamaks
field aligned plasma pressure perturbations universally appear. These perturbations
are spatially localized when viewed in a plane perpendicular to the magnetic field and
are often referred to as blobs. They mediate a significant amount of the radial
particle and energy flux on plasma facing components and thus critically
determine their lifetime~\cite{Krasheninnikov2001, Antar2001, DIppolito2002, Boedo2003,
Garcia2005, Garcia_Bian_Fundamensky_POP_2006, Myra2006, Theiler2009, Carralero2015}.
Recent efforts in stochastic modeling relate the radial density profiles of
magnetically confined plasmas to the amplitude, size and radial velocity of
individual uncorrelated transport events such as blobs~\cite{Garcia2016}.  Analysis of
experimental data support the predictions of this stochastic model: probability
density functions, auto correlation and power spectra as well as threshold
level crossings of the turbulent fields are in good agreement with theoretical predictions~\cite{Garcia2012,
Garcia2013, Garcia2015, Theodorsen2016, Kube2016a, Garcia2016}.

A similar transport mechanism is believed to act in the F-layer ionosphere.
Here depletions in the plasma density or ``bubbles'' are observed in night-side
equatorial regions.
The rising plasma depletions are thought to trigger turbulent flows in
otherwise stable regions and lead to the equatorial spread-F phenomenon, which 
may significantly affect the performance and reliability of radio frequency
transmissions~\cite{Cohen1961, Woodman1976, Ott1978, Hysell1998, Hysell2000,
Hysell2002, Woodman2009}.
Measurements of plasma depletions have also been reported from magnetically
confined plasmas although their contribution to transport of plasma is still
debated~\cite{Boedo2003, Cheng2010, Nold2010}.  

In this contribution scrape-off layer plasmas as well as ionospheric plasmas
are modeled by drift-fluid equations where we ignore magnetic field inhomogenity
for the latter one. This simplification results in incompressible flows.
As noted in~\cite{Kube2016}, compressible drifts significantly alter the 
dynamics of seeded perturbations with low peak amplitudes relative to the 
background level. We further discuss the effect of the seeded perturbations'
inertial mass on the acceleration of the structure~\cite{Kendl2015}. Using
the conservation laws of the model equations we derive an expression that
relates the acceleration of pressure perturbations to its initial amplitude
relative to the background. An empirical model is proposed that is shown
to reproduce velocities and accelerations taken from numerical simulations
over a broad range of initial density amplitudes.


In drift-fluid models the continuity equation
\begin{align}
 \frac{\partial n}{\partial t} + \nabla\cdot\left( n \vec u_E  \right) &= 0 \label{eq:generala} 
\end{align}
describes the dynamics of the electron density $n$. Here
$\vec u_E := (\bhat \times \nabla \phi)/B$ gives the electric drift
velocity in a magnetic field $\vec B := B \bhat$ and an electric
potential $\phi$. We neglect contributions of the diamagnetic drift~\cite{Kube2016}.


Equation~\eqref{eq:generala} is closed by invoking quasineutrality, i.e. the divergence of the ion polarization, 
the electron diamagnetic and the gravitational drift currents must vanish
\begin{align}
  \nabla\cdot\left( \frac{n}{\Omega} \left( \frac{\partial}{\partial t} 
  + \vec u_E \cdot\nabla  \right)\frac{\nabla_\perp \phi}{B}  + n\vec u_d - n\vec u_g\right) &=0
  . 
  \label{eq:generalb}
\end{align}
Here we denote 
$\nabla_\perp\phi/B := - \bhat \times \vec u_E$, 
the electron diamagnetic drift
$\vec u_d := - T_e(\bhat \times\nabla n ) /enB$
with the electron temperature $T_e$,
the ion gravitational drift velocity  
$\vec u_g := m_i \bhat \times \vec g /B$
with ion mass $m_i$, and the ion gyro-frequency
$\Omega := eB/m_i$.

Combining Eq.~\eqref{eq:generalb} with Eq.~\eqref{eq:generala} yields
\begin{align}
 \frac{\partial \rho}{\partial t} + \nabla\cdot\left( \rho\vec u_E \right) + \nabla \cdot\left( n(\vec u_\psi + \vec u_d + \vec u_g) \right) &= 0\label{eq:vorticity}
\end{align}
with the polarization charge density 
$\rho = \nabla\cdot( n\nabla_\perp \phi / \Omega B)$ 
and
$\vec u_\psi := \bhat\times \nabla\psi /B$ 
with 
$\psi:= m_i\vec u_E^2 /2e$.
We exploit this form of Eq.~\eqref{eq:generalb} in our numerical simulations.

Equations~\eqref{eq:generala} and \eqref{eq:generalb} respectively \eqref{eq:vorticity} have several invariants.
First, in Eq.~\eqref{eq:generala} the relative particle number 
$M(t) := \int \dV (n-n_0)$ is conserved over time
$\d M(t)/\d t = 0$. 
Furthermore, we integrate 
$( T_e(1+\ln n) -T_e \ln B)\partial_t n$
as well as
$-e\phi \partial_t\rho - (m_i\vec u_E^2/2+gm_ix - T_e\ln B)\partial_t n$ 
over the domain to get, disregarding boundary contributions,
\begin{align}
  \frac{\d}{\d t}\left[T_eS(t) + H(t) \right] = 0, \label{eq:energya}\\ 
    \frac{\d}{\d t} \left[ E(t) - G(t) - H(t)\right] =  0,
    \label{eq:energyb}
\end{align}
where we define 
the entropy
$S(t):=\int \dV [n\ln(n/n_0) - (n-n_0)]$,  
the kinetic energy 
$E(t):=m_i \int \dV n\vec u_E^2/2$ 
and the potential energies
$G(t) := m_i g\int \dV x(n-n_0)$
and
$H(t) := T_e\int \dV (n-n_0) \ln (B^{-1})$.
Note that $n\ln( n/n_0) - n + n_0 \approx (n-n_0)^2/2$ for $|(n-n_0)/n_0| \ll 1$ and $S(t)$ thus reduces to the 
local entropy form in Reference~\cite{Kube2016}. 

We now set up a gravitational field $\vec g = g\hat x$ and a constant homogeneous background
magnetic field $\vec B = B_0 \hat z$ in a Cartesian coordinate system.
Then the divergences of the electric and gravitational drift velocities $\nabla\cdot\vec u_E$ and $\nabla\cdot\vec u_g$
and the diamagnetic current $\nabla\cdot(n\vec u_d)$ vanish, which makes the 
flow incompressible. Furthermore, the magnetic potential energy vanishes $H(t) = 0$.

In a second system we model the inhomogeneous magnetic field present in tokamaks as
$\vec B := B_0 (1+ x/R_0)^{-1}\hat z$ and neglect the gravitational drift $\vec u_g = 0$.
Then, the potential energy $G(t) = 0$. 
Note that 
$H(t) = m_i \Cs^2/R_0\int\dV x(n-n_0) +\mathcal O(R_0^{-2}) $
reduces to $G(t)$ with the effective gravity $g_\text{eff}:= \Cs^2/R_0$ with $\Cs^2 := T_e/m_i$. 
For the rest of this letter we treat $g$ and $g_\text{eff}$ as well as $G(t)$ and $H(t)$ on the same footing.
The magnetic field inhomogeneity thus entails compressible flows, which is 
the only difference to the model describing dynamics in a homogeneous magnetic field introduced above. 
Since both $S(t)\geq 0$ and $E(t)\geq 0$ we further derive from Eq.~\eqref{eq:energya} and Eq.~\eqref{eq:energyb} that the kinetic energy
is bounded by $E(t) \leq T_eS(t) + E(t) = T_e S(0)$; a feature absent from the gravitational system with 
incompressible flows, where $S(t) = S(0)$. 

We now show that the invariants Eqs.~\eqref{eq:energya} and \eqref{eq:energyb} present restrictions on the velocity and
acceleration of plasma blobs. 
First, we define the blobs' center of mass (COM) via $X(t):= \int\dV x(n-n_0)/M$ and 
its COM velocity as $V(t):=\d X(t)/\d t$. 
The latter is proportional to the total radial particle flux~\cite{Garcia_Bian_Fundamensky_POP_2006, Held2016a}.
We assume
that $n>n_0$ and $(n-n_0)^2/2 \leq [ n\ln (n/n_0) - (n-n_0)]n $ to show for both systems 
\begin{align}
  (MV)^2 &= \left( \int \dV n{\phi_y}/{B} \right)^2
  = \left( \int \dV (n-n_0){\phi_y}/{B} \right)^2\nonumber\\
&\leq 2 \left( \int \dV \left[n\ln (n/n_0) -(n-n_0)\right]^{1/2}\sqrt{n}{\phi_y}/{B}\right)^2\nonumber\\
  &\leq 4 S(0) E(t)/m_i 
  \label{eq:inequality}
\end{align}
Here we use the Cauchy-Schwartz inequality and 
$\phi_y:=\partial\phi/\partial y$. 
Note that although we derive the inequality Eq.~\eqref{eq:inequality} only for amplitudes $\triangle n >0$  we assume that the results also hold for depletions. This is justified by our numerical results later in this letter. 
If we initialize our density field with a seeded blob of radius $\ell$ and amplitude $\triangle n$ as 
\begin{align}
  n(\vec x, 0) &= n_0 + \triangle n \exp\left( -\frac{\vec x^2}{2\ell^2} \right), \label{eq:inita}
\end{align}
and  
$\phi(\vec x, 0 ) = 0$,
we immediately have $M := M(0) = 2\pi \ell^2 \triangle n$, $E(0) = G(0) = 0$ and 
$S(0) = 2\pi \ell^2 f(\triangle n)$, where $f(\triangle n)$ captures the amplitude dependence of 
the integral for $S(0)$. 

The acceleration for both incompressible and compressible flows can be estimated
by assuming a linear acceleration $V=A_0t$ and $X=A_0t^2/2$~\cite{Held2016a} and using 
$E(t) = G(t) = m_igMX(t)$ in Eq.~\eqref{eq:inequality} 
\begin{align}
  \frac{A_0}{g} =  \mathcal Q\frac{2S(0)}{M} \approx \frac{\mathcal Q}{2} \frac{\triangle n }{n_0+2\triangle n/9}.
  \label{eq:acceleration}
\end{align}
Here, we use the Pad\'e approximation of order $(1/1)$ of $2S(0)/M $
and define a model parameter $\mathcal Q$ with $0<\mathcal Q\leq1$ to be determined by numerical simulations.
Note that the Pad\'e approximation is a better approximation than a simple 
truncated Taylor expansion especially for large relative amplitudes of order unity.
Eq.~\eqref{eq:acceleration} predicts that $A_0/g\sim \triangle n/n_0$ for small 
amplitudes $|\triangle n/n_0| < 1$ and $A_0 \sim g $ for very large amplitudes $\triangle n /n_0 \gg 1$, 
which confirms the predictions in~\cite{Pecseli2016} and reproduces the limits discussed in~\cite{Angus2014}.

As pointed out earlier for compressible flows Eq.~\eqref{eq:inequality} can be further estimated
\begin{align}
  (MV)^2  \leq 4 T_eS(0)^2/m_i. 
  \label{}
\end{align}
We therefore have a restriction on the maximum COM velocity for compressible flows, which is absent for incompressible flows
\begin{align}
  \frac{\max |V|}{\Cs} = {\mathcal Q}\frac{2S(0)}{M} \approx \frac{\mathcal Q}{2} \frac{|\triangle n| }{n_0+2/9 \triangle n } \approx \frac{\mathcal Q}{2} \frac{|\triangle n|}{n_0}.
  \label{eq:linear}
\end{align}
For $|\triangle n /n_0|< 1$ Eq.~\eqref{eq:linear} reduces to the linear scaling derived in~\cite{Kube2016}. 
Finally, a scale analysis of Eq.~\eqref{eq:vorticity} shows that~\cite{Ott1978, Garcia2005, Held2016a}
\begin{align}
  \frac{\max |V|}{\Cs} = \mathcal R \left( \frac{\ell}{R_0}\frac{|\triangle n|}{n_0} \right)^{1/2}.
  \label{eq:sqrt}
\end{align}
This equation predicts a square root dependence of the center of mass velocity 
on amplitude and size. 


We now propose a simple phenomenological model that captures the essential dynamics
of blobs and depletions in the previously stated systems. More specifically 
the model reproduces the acceleration Eq.~\eqref{eq:acceleration} with and without
Boussinesq approximation, the square root scaling for the COM velocity 
Eq.~\eqref{eq:sqrt} for incompressible flows as well as the relation between the 
square root scaling Eq.~\eqref{eq:sqrt} and the linear scaling 
Eq.~\eqref{eq:linear} for compressible flows. 
The basic idea is that the COM of blobs behaves like 
the one of an infinitely long plasma column immersed in an ambient plasma. 
The dynamics of this column reduces to the one of a two-dimensional ball.
This idea is similar to the analytical ``top hat'' density solution for
blob dynamics recently studied in~\cite{Pecseli2016}.
The ball is subject to buoyancy as well as linear and nonlinear friction
\begin{align}
  M_{\text{i}} \frac{d V}{d t} = (M_{\text{g}} - M_\text{p}) g - c_1 V  - \mathrm{sgn}(V ) \frac{1}{2}c_2 V^2.
  \label{eq:ball}
\end{align}
The gravity $g$ has a positive sign in the coordinate system; sgn$(f)$ is the sign function. 
The first term on the right hand side is the buoyancy, where 
$M_{\text{g}} := \pi \ell^2 (n_0 + \mathcal Q \triangle n/2)$ 
is the gravitational mass of the ball with radius $\ell$ and 
$M_\mathrm{p} := n_0 \pi \ell^2 $ 
is the mass of the displaced ambient plasma.
Note that if $\triangle n<0$ the ball represents a depletion and the buoyancy term has a negative sign, i.e. the depletion will rise. 
We introduce an inertial mass 
$M_{\text{i}} := \pi\ell^2 (n_0 +2\triangle n/9)$ 
different from the gravitational mass $M_{\text{g}}$ in order to 
recover the initial acceleration in Eq.~\eqref{eq:acceleration}. 
We interpret the parameters $\mathcal Q$ and $2/9$ as geometrical factors 
that capture the difference of the actual blob form from the idealized
``top hat'' solution. 
Also note that the Boussinesq approximation appears in the model as a neglect of inertia, $M_{\text{i}} = \pi\ell^2n_0$.

The second term is the linear friction term with coefficient $c_1(\ell)$, which
depends on the size of the ball.
If we disregard the nonlinear friction, $c_2=0$, Eq.~\eqref{eq:ball} directly yields a 
maximum velocity $c_1V^*=\pi \ell^2 n g \mathcal Q\triangle n/2$.
From our previous considerations $\max V/\Cs=\mathcal Q \triangle n /2n_0$, we thus identify 
\begin{align}
  c_1 = \pi\ell^2 n_0 g/\Cs.  
  \label{}
\end{align}
The linear friction coefficient thus depends on the gravity and the size of the
ball. 

The last term in \eqref{eq:ball} is the nonlinear friction. The sign of the force depends on whether
the ball rises or falls in the ambient plasma. 
If we disregard linear friction $c_1=0$, we have the maximum velocity 
$V^*= \sigma(\triangle n)\sqrt{\pi \ell^2|\triangle n| g\mathcal Q/c_2}$, 
which must equal 
$\max V= \sigma(\triangle n) \mathcal R \sqrt{g \ell |\triangle n/n_0|}$ 
and thus
\begin{align}
  c_2 = {\mathcal Q\pi n_0\ell }/{\mathcal R^2}.
  \label{}
\end{align}
%
Inserting $c_1$ and $c_2$ into Eq.~\eqref{eq:ball}
we can derive the maximum absolute velocity in the form 
\begin{align}
  \frac{\max |V|}{\Cs} = 
        \left(\frac{\mathcal R^2}{\mathcal Q}\right) \frac{\ell}{R_0} \left( 
        \left({1+\left( \frac{\mathcal Q}{\mathcal R} \right)^{2} \frac{|\triangle n|/n_0 }{\ell/R_0}}\right)^{1/2}-1 \right)
  \label{eq:vmax_theo}
\end{align}
and thus have a concise expression for $\max |V|$ that captures both the linear
scaling \eqref{eq:linear} as well as the square root scaling \eqref{eq:sqrt}.
With Eq.~\eqref{eq:acceleration} and Eq.~\eqref{eq:sqrt} respectively Eq.~\eqref{eq:vmax_theo} we 
finally arrive at an analytical expression for the time at which the maximum velocity is reached via 
$t_{\max V} \sim \max V/A_0$. Its inverse $\gamma:=t_{\max V}^{-1}$ gives the
global interchange growth rate, for which an empirical expression was
presented in Reference~\cite{Held2016a}.

We use the open source library FELTOR 
to simulate 
Eqs.~\eqref{eq:generala} and \eqref{eq:vorticity} with and without 
drift compression.
For numerical stabilty we added small diffusive terms on the right hand 
sides of the equations.
The discontinuous Galerkin methods employ three polynomial coefficients and a minimum of $N_x=N_y=768$ grid cells. The box size is $50\ell$ in order to mitigate 
influences of the finite box size on the blob dynamics. 
Moreover, we used the invariants in Eqs. \eqref{eq:energya} and \eqref{eq:energyb} as consistency tests to verify the code and repeated simulations 
also in a gyrofluid model. 
No differences to the results presented here were found. 
Initial perturbations on the particle density field are given by Eq.~\eqref{eq:inita},
where the perturbation amplitude $\triangle n/n_0$ was chosen between $10^{-3}$ and $20$ for blobs and $-10^0$ and $ -10^{-3}$ for depletions. 
Due to computational reasons we show results only for $\triangle n/n_0\leq 20$. 
For compressible flows we consider two different cases $\ell/R_0 = 10^{-2}$ and
$\ell /R_0 = 10^{-3}$. 
 For incompressible flows Eq.~\eqref{eq:generala} and \eqref{eq:vorticity}
 can be normalized such that the blob radius is absent from the equations~\cite{Ott1978, Kube2012}. 
 The simulations of incompressible flows can thus be used for both sizes. 
The numerical code as well as input parameters and output data can be found 
in the supplemental dataset to this contribution~\cite{Data2017}.

\begin{figure}[htb]
    \includegraphics[width=\columnwidth]{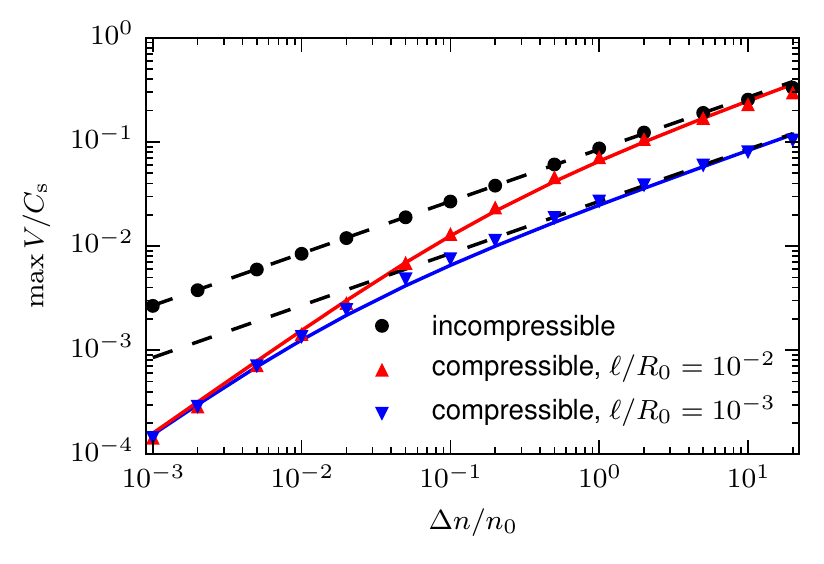}
    \caption{
      The maximum radial COM velocities of blobs for compressible and incompressible flows are shown. 
      The continuous lines show Eq.~\eqref{eq:vmax_theo} while the 
      dashed line shows the square root scaling Eq.~\eqref{eq:sqrt} with 
      $\mathcal Q = 0.32$ and $\mathcal R=0.85$.
    }
    \label{fig:com_blobs}
\end{figure}
In Fig.~\ref{fig:com_blobs} we plot the maximum COM velocity for blobs 
with and without drift compression.
For incompressible flows blobs follow the square root scaling almost 
perfectly. Only at very large amplitudes velocities are slightly below
the predicted values. 
For small amplitudes we observe that the compressible blobs follow
a linear scaling. When the amplitudes increase there is a transition to the
square root scaling at around $\triangle n/n_0 \simeq 0.5$ for 
$\ell/R_0=10^{-2}$ and $\triangle n/n_0 \simeq 0.05$ for $\ell/R_0=10^{-3}$, which is consistent with Eq.~\eqref{eq:vmax_theo} and Reference~\cite{Kube2016}. 
In the transition regions the simulated velocities are slightly larger than the predicted ones from Eq.~\eqref{eq:vmax_theo}.
Beyond these amplitudes
the velocities of compressible and incompressible blobs align. 

\begin{figure}[htb]
    \includegraphics[width=\columnwidth]{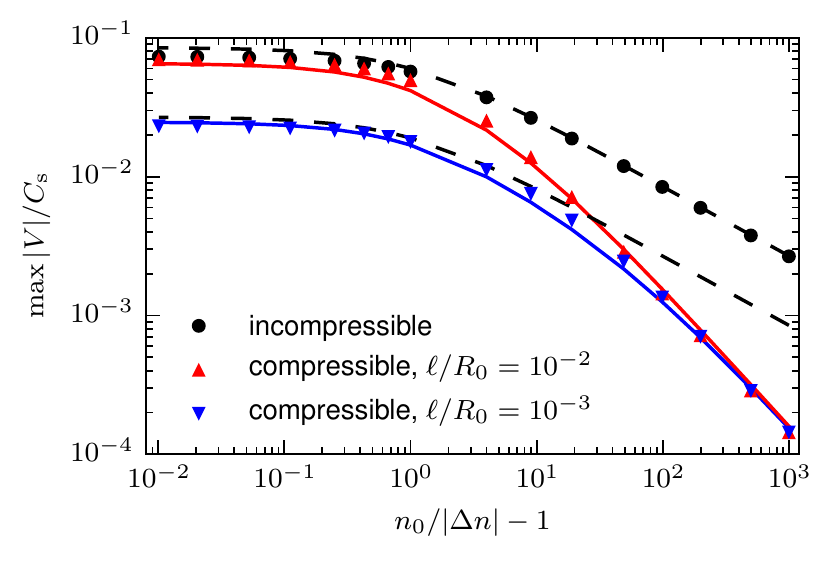}
    \caption{
      The maximum radial COM velocities of depletions for compressible and incompressible flows are shown. 
      The continuous lines show Eq.~\eqref{eq:vmax_theo} while the 
      dashed line shows the square root scaling Eq.~\eqref{eq:sqrt} with 
      $\mathcal Q = 0.32$ and $\mathcal R=0.85$.
      Note that small amplitudes are on the right and amplitudes close to unity are on the left side.
  }
    \label{fig:com_depletions}
\end{figure}
In Fig.~\ref{fig:com_depletions} we show the maximum radial COM velocity 
for depletions instead of blobs.
For relative amplitudes below $|\triangle n|/n_0 \simeq 0.5$ (right of unity in the plot) the velocities
coincide with the corresponding blob velocities in Fig.~\ref{fig:com_blobs}. 
 For amplitudes larger than $|\triangle n|/n_0\simeq 0.5$ the 
velocities follow the square root scaling.
We observe that for plasma depletions beyond $90$ percent the velocities 
in both systems reach a constant value that is very well predicted by the
square root scaling. 

\begin{figure}[htb]
    \includegraphics[width=\columnwidth]{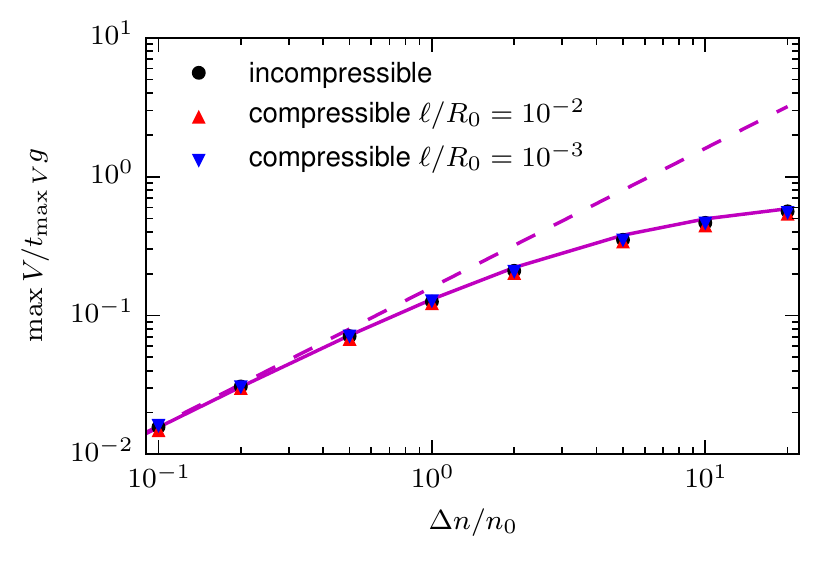}
    \caption{
      Average acceleration of blobs for compressible and incompressible flows are shown.
      The continuous line shows the acceleration in Eq.~\eqref{eq:acceleration} 
      with $\mathcal Q=0.32$
      while the dashed line is a linear reference line, which corresponds to the Boussinesq approximation. 
  }
    \label{fig:acc_blobs}
\end{figure}
In Fig.~\ref{fig:acc_blobs} we show the average acceleration of blobs 
for compressible and incompressible flows computed
by dividing the maximum velocity $\max V$ by the time  
to reach this velocity $t_{\max V}$. 
We compare the simulation results
to the theoretical predictions Eq.~\eqref{eq:acceleration} of our model with and without inertia. 
The results of the compressible and incompressible systems coincide and fit very
well to our theoretical values. 
For amplitudes larger than unity the acceleration deviates significantly from the prediction with Boussinesq approximation.

\begin{figure}[htb]
    \includegraphics[width=\columnwidth]{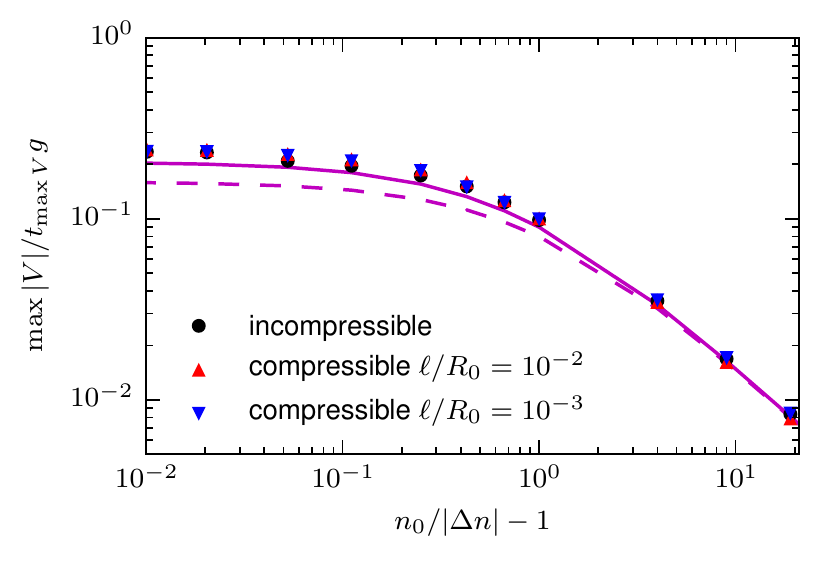}
    \caption{
      Average acceleration of depletions for compressible and incompressible flows are shown.
      The continuous line shows the acceleration in Eq.~\eqref{eq:acceleration} 
      with $\mathcal Q=0.32$
      while the dashed line is a linear reference line, which corresponds to the Boussinesq approximation. 
    }
    \label{fig:acc_depletions}
\end{figure}
In Fig.~\ref{fig:acc_depletions} we show the simulated acceleration of depletions in the
compressible and the incompressible systems. We compare the simulation results
to the theoretical predictions Eq.~\eqref{eq:acceleration} of our model with and without inertia.
Deviations from our theoretical prediction Eq.~\eqref{eq:acceleration} are visible for amplitudes smaller than $\triangle n/n_0 \simeq -0.5$ (left of unity in the plot). The relative deviations are small at around $20$ percent. 
As in Fig.~\ref{fig:com_depletions} the acceleration reaches a constant values
for plasma depletions of more than $90$ percent.
Comparing Fig.~\ref{fig:acc_depletions} to Fig.~\ref{fig:acc_blobs} the asymmetry between blobs and depletions becomes 
apparent. While the acceleration of blobs is reduced for large 
amplitudes compared to a linear dependence the acceleration 
of depletions is increased. In the language of our simple buoyancy 
model the inertia of depletions is reduced but increased for blobs.

In conclusion  
  we discuss the dynamics of seeded blobs and depletions in a 
  compressible and an incompressible system.
  With only two fit parameters our theoretical results reproduce the 
  numerical COM velocities and accelerations over five orders of magnitude.
  We derive the amplitude dependence of the acceleration of blobs and depletions from 
  the conservation laws of our systems in Eq.~\eqref{eq:acceleration}. 
  From the same inequality a linear regime is derived in the compressible system for 
  ratios of amplitudes to sizes smaller than a critical value.
   In this regime 
  the blob and depletion velocity depends linearly on the initial amplitude and 
  is independent of size. The regime is absent from the system with incompressible flows.
  Our theoretical results are verified by numerical simulations for all 
  amplitudes that are relevant in magnetic fusion devices.
  Finally, we suggest a new empirical blob model that captures the detailed dynamics of more complicated models. 
  The Boussinesq approximation is clarified as the absence of inertia and a thus altered acceleration of blobs and depletions.
  The maximum blob velocity is not altered by the Boussinesq approximation.

The authors were supported with financial subvention from the Research Council of Norway under grant
240510/F20. M.W. and M.H. were supported by the Austrian Science Fund (FWF) Y398.  The computational
results presented have been achieved in part using the Vienna Scientific Cluster (VSC). Part of this work was performed on the Abel Cluster, owned by the University of Oslo and the Norwegian metacenter
for High Performance Computing (NOTUR), and operated by the Department for Research Computing at USIT,
the University of Oslo IT-department.
This work has been carried out within the framework of the EUROfusion Consortium and has received funding from the Euratom research and training programme 2014-2018 under grant agreement No 633053. The views and opinions expressed herein do not necessarily reflect those of the European Commission.


\bibliography{refs.bib}

\end{document}